\documentclass[aps,prb,twocolumn,a4,showpacs,superscriptaddress]{revtex4-1} 
\usepackage{amsmath,amssymb,amsfonts,upgreek}
\usepackage{mathtools}

\usepackage[english]{babel}
\usepackage[utf8x]{inputenc}
\usepackage[T1]{fontenc}
\usepackage{bm}
\usepackage{nicefrac}
\usepackage{siunitx}
\usepackage{textcomp}
\usepackage{booktabs}

\usepackage[figuresright]{rotating}

\usepackage[colorinlistoftodos]{todonotes}
\usepackage[colorlinks=true, allcolors=blue]{hyperref}
\usepackage{relsize}
\usepackage{amsmath,amssymb,amsfonts,upgreek}
\usepackage{gensymb}
\usepackage{color}
\usepackage{graphicx}
\usepackage{setspace}
\usepackage{multirow}
\usepackage{hhline}
\usepackage{bm}
\usepackage{comment}
\usepackage{natbib}

\begin{document}
\title{First-principles Landau-like potential for BiFeO$_3$ and related materials}

\author{Natalya S.\ Fedorova}
\email{natalya.fedorova@list.lu}
\affiliation{Materials Research and Technology Department, Luxembourg Institute of Science and Technology,
5 Avenue des Hauts-Fourneaux, L-4362 Esch/Alzette, Luxembourg}

\author{Dmitri E. Nikonov}
\affiliation{Components Research, Intel Corporation, Hillsboro, 97124 Oregon, USA}

\author{Hai Li}
\affiliation{Components Research, Intel Corporation, Hillsboro, 97124 Oregon, USA}

\author{Ian A. Young}
\affiliation{Components Research, Intel Corporation, Hillsboro, 97124 Oregon, USA}

\author{Jorge \'{I}\~{n}iguez}
\email{jorge.iniguez@list.lu}
\affiliation{Materials Research and Technology Department, Luxembourg Institute of Science and Technology,
5 Avenue des Hauts-Fourneaux, L-4362 Esch/Alzette, Luxembourg}
\affiliation{Department of Physics and Materials Science, University of Luxembourg, 41 Rue du Brill, L-4422 Belvaux, Luxembourg}

\begin{abstract}
In this work we introduce the simplest, lowest-order Landau-like
potential for BiFeO$_3$ and La-doped BiFeO$_3$ as an expansion around
the paraelectric cubic phase in powers of polarization, FeO$_6$
octahedral rotations and strains. We present an analytical approach
for computing the model parameters from density functional theory. We
illustrate our approach by computing the potentials for BiFeO$_3$ and
La$_{0.25}$Bi$_{0.75}$FeO$_3$ and show that, overall, we are able to
capture the first-principles results accurately. The computed models
allow us to identify and explain the main interactions controlling the
relative stability of the competing low-energy phases of these
compounds.
\end{abstract}


\maketitle

\section{Introduction}
\label{sec:introduction}

Magnetoelectric multiferroics, materials that simultaneously show magnetic and
electric orders, are of significant interest, since the coexistence
and coupling of these orders hold great potential for development of
multifunctional devices \cite{spaldin2005,eerenstein2006}. BiFeO$_3$
is among the most exciting and extensively studied representatives of
this family because it displays both orders at room temperature
\cite{catalan2009}.

Ferroelectricity appears in BiFeO$_3$ at $T_C \sim 1100$~K
\cite{moreau1971,smith1968}. Below $T_C$, it has a rhombohedrally
distorted perovskite structure (space group $R3c$, \#161)
\cite{michel1969,kubel1990}, which differs from the perfect cubic
phase by the presence of two distortions: (i) polar displacements of
Bi$^{3+}$ and Fe$^{3+}$ cations with respect to O$^{2-}$ anions
(Bi$^{3+}$ dominates due to its stereochemically active $6s$ lone
pairs \cite{seshadri2001}) giving rise to a spontaneous polarization
$\mathbf{P}$ of up to 100 $\mu$C/cm$^2$ along a pseudocubic $\langle
111\rangle$ direction \cite{lebeugle2007_2,wang2003}; and (ii)
antiphase rotations $\mathbf{R}$ of the FeO$_6$ octahedra about the same
pseudocubic $\langle 111\rangle$ direction as the polarization ($a^-a^-a^-$ in Glazer's
notation \cite{glazer1972}) \cite{ederer2005,dieguez2011}. (In the
following, all directions are in the pseudocubic setting.)

Below $T_N \sim 640$~K, BiFeO$_3$ also shows G-type antiferromagnetic
(G-AFM) order with the nearest-neighboring Fe spins antialigned
\cite{bhide1965,moreau1971}. The canting of the Fe spins driven by
Dzyaloshinskii-Moriya (DM) interaction
\cite{dzyaloshinsky1958,moriya1960} can give rise to a weak
magnetization in this material. The DM interaction relies on the
symmetry breaking caused by the FeO$_6$ octahedral tilts of
BiFeO$_{3}$ \cite{ederer2005}; indeed, the phase of the octahedral
rotations defines the sign of the DM vector and, in turn, that of the
weak magnetization. In bulk BiFeO$_3$ an incommensurate cycloidal
spiral is superimposed on the G-AFM order, yielding a zero net magnetization
\cite{sosnowska1982}. This cycloid, however, can be suppressed by
doping in bulk systems \cite{sosnowska2002} and by epitaxial
constraints in BiFeO$_3$ films
\cite{bai2005,bea2007,sando2013,heron2014}.  Therefore,
ferroelectricity can coexist with weak ferromagnetism in BiFeO$_3$ at
ambient conditions.

Additionally, a 180$^{\circ}$ deterministic switching of the DM vector
and weak magnetization by an electric field has been reported from a
combined experimental and theoretical study of BiFeO$_3$ films grown
on DyScO$_{3}$ substrates \cite{heron2014}. It is proposed that the
magnetoelectric switching is the result of a peculiar polarization
reversal that is found to occur in two steps, a 109$^{\circ}$ rotation
followed by a 71$^{\circ}$ rotation (or \textit{vice versa}); further,
the FeO$_6$ octahedral tilts are believed to reverse together with the
polarization, resulting in the observed reversal of the weak magnetic
moment. Note that octahedral tilts will typically not follow
polarization in a single-step 180$^{\circ}$ reversal and, therefore, a
two-step switching path is crucial for controlling the weak
magnetization in BiFeO$_3$ by an electric field. These observations
make BiFeO$_3$ a promising candidate for applications in
magnetoelectric memory elements. However, to be technologically
relevant, switching characteristics have to be optimized such that
coercive voltages are below 100~mV and switching times fall in the
range of 10-1000~ps \cite{manipatruni2018,prasad2020}. Hence, the
current challenge is to optimize the ferroelectric switching in
BiFeO$_{3}$ while retaining the two-step path and magnetoelectric
control.

One of the efficient strategies for optimizing polarization switching
in BiFeO$_3$ is doping by La.  Indeed, since polarization in this
compound largely originates from the $6s$ lone pairs
of the Bi$^{3+}$ cations, their substitution by isovalent, lone-pair-free
cations leads to a reduction of the polar distortion
\cite{catalan2009,chu2008,gonzalez2012}. For example, it has been experimentally
demonstrated that 15-20\% La-doped BiFeO$_3$ films show a polarization
which is up to 60\% smaller than that of pure BiFeO$_3$ films
\cite{prasad2020,zhang2019}. Further, first-principles calculations
have predicted that subsitution of Bi by La cations reduces the energy
barrier between polar states by up to 50\% for 25\% doping. This, in
turn, leads to a reduction of coercive voltages (down to 0.8~V for a
100~nm film), enabling low-power switching
\cite{prasad2020}. Additionally, a significant reduction of switching
times has been demonstrated for La$_{0.15}$Bi$_{0.75}$FeO$_3$ films
compared to pure BiFeO$_3$ in a wide range of applied electric fields
\cite{parsonnet2020}. Nevertheless, further improvement requires
understanding the origin of the two-step polarization switching in
BiFeO$_3$ and related materials, as well as search for other
strategies for manipulating the switching energy landscape. For that
purpose, dynamical simulations of polarization switching based on
phenomenological models of the free energy can be very helpful.

Landau free-energy potentials
\cite{landau1937,landau1937_2,devonshire1949,devonshire1951}, together
with the Landau-Khalatnikov time-evolution equation \cite{umantsev2012}, offer a
practical scheme to investigate switching in ferroelectrics. In this approach, one expands the energy of the
compound around the reference paraelectric phase in
powers of the relevant order parameters, keeping only terms compatible
with the crystal symmetry \cite{rabe2007}. It is important to note
that the reliability of such simulations depends on the choice of the
free energy expansion's coefficients, which can be obtained either by
fitting to experimental data or from first-principles calculations
\cite{rabe2007}. In compounds as complex as BiFeO$_{3}$, which feature
multiple primary order parameters, deriving a suitable Landau
potential from experimental information is all but impossible; hence,
there is a clear need for the development of first-principles
approaches.

In this work we introduce the simplest, lowest-order Landau-like
potential able to reproduce the energies and structures of the
low-energy polymorphs of BiFeO$_3$ and related materials. We present
an analytical approach to compute the model parameters from density
functional theory (DFT) and apply it to BiFeO$_{3}$ and
La$_{0.25}$Bi$_{0.75}$FeO$_{3}$. We demonstrate the overall accuracy
of the obtained potentials, and discuss an effective way to treat
intermediate compositions. Finally, we discuss the physics captured by
the model, namely, the interaction between polarization and FeO$_6$
octahedral tilts, how it affects the energetics of different BiFeO$_3$
polymorphs, as well as the effects of La doping.

\begin{figure}
    \centering
    \includegraphics[width=0.98\linewidth,trim=0cm 0cm 0cm 0cm]{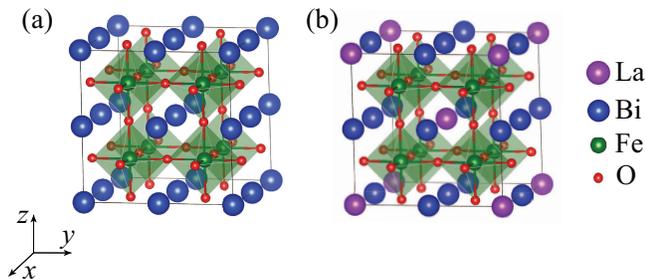}
    \caption{Sketch of the 40-atom supercell used in our simulations of (a) BiFeO$_3$ and (b) La$_{0.25}$Bi$_{0.75}$FeO$_3$.}
    \label{fig:crystal_structure}
\end{figure}

\begin{figure}
    \centering
    \includegraphics[width=0.92\linewidth,trim=0cm 0cm 0cm 0cm]{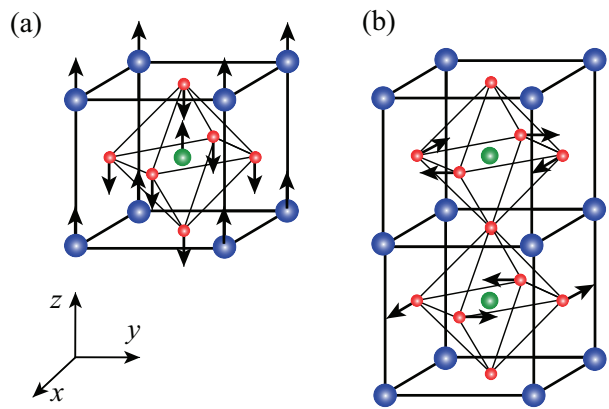}
    \caption{Ionic displacement patterns corresponding to (a) a polar distortion mode along the [001] pseudocubic direction and (b) FeO$_6$ octahedral rotations about the same axis. Arrows indicate the directions of the ionic displacements and do not reflect their relative amplitudes. Blue, green and red circles indicate Bi, Fe and O ions, respectively. }
    \label{fig:distortion_modes}
\end{figure}


\section{Computational details}
\label{sec:comp_details}

All calculations are performed using the DFT
\cite{hohenberg1964,kohn1965} implementation in the Vienna \textit{Ab
  initio} Simulation package (VASP) \cite{kresse1996}. For the
exchange-correlation potential, we use the generalized gradient
approximation optimized for solids \cite{perdew2008}, with a Hubbard
$U$ correction (within Dudarev's scheme \cite{dudarev1998} and
$U=4$~eV) for a better treatment of iron's 3$d$ electrons. We treat
the interaction between core and valence electrons by the
projector-augmented plane wave method \cite{blochl1994,kresse1996},
solving explicitly for 15 electrons of Bi ($5d^{10}6s^26p^3$), 9 of La
($5p^66s^25d^1$), 14 of Fe ($3p^63d^74s^1$), and 6 of O
($2s^22p^4$). We use a plane-wave basis set with a cutoff energy of
500~eV. We use a $3\times3\times3$ $\Gamma$-centered Monkhorst-Pack
$k$-point grid for reciprocal space integrals in the Brillouin zone
corresponding to a 40-atom cell that is a 2$\times$2$\times$2 multiple
of the 5-atom perovskite unit (see
Fig.~\ref{fig:crystal_structure}(a)). We ensure that these choices provide a good level of convergence for the
quantities of interest. All simulations are performed with the G-type
antiferromagnetic order of Fe magnetic moments imposed. In the lattice
optimizations, the structures are considered to be relaxed when the
forces acting on the atoms are below 0.01~eV/\AA. We calculate elastic
constants by finite differences using the strain-stress relationship
\cite{lepage2002}.


\section{Formalism}
\label{sec:formalism}
\subsection{Landau-like potential}
\label{subsec:LD_potential}

In this section we introduce the potential for BiFeO$_3$ and La-doped
BiFeO$_3$ as an expansion around the reference paraelectric cubic
phase in powers of the following order parameters: (i) the
three-dimensional electric polarization $\mathbf{P}=(P_x,P_y,P_z)$;
(ii) the antiphase rotations of FeO$_6$ octahedra
$\mathbf{R}=(R_x,R_y,R_z)$; (iii) the strain $\bm{\eta}
=(\eta_{xx},\eta_{yy},\eta_{zz},\eta_{yz},\eta_{xz},\eta_{xy})$, where
$\eta_{xx}=\epsilon_{xx}$, $\eta_{yy}=\epsilon_{yy}$,
$\eta_{zz}=\epsilon_{zz}$, $\eta_{yz}=2\epsilon_{yz}$,
$\eta_{xz}=2\epsilon_{xz}$ and $\eta_{xy}=2\epsilon_{xy}$, and
$\epsilon_{ij}$ are the components of the homogeneous strain
tensor. The resulting expression for the potential (per perovskite
unit cell) is written as follows:
\begin{equation}
\label{eq:F_full}
\begin{aligned}
    F(P,R,\eta)=&F_0 + F(P)+F(R)+F(\eta)+ \\ 
    & F(P,R)+F(P,\eta)+F(R,\eta).
    \end{aligned}
\end{equation}
Here, $F_0$ is the free energy of the reference cubic phase. $F(P)$,
$F(R)$ and $F(\eta)$ are the energy contributions solely due to
polarization, FeO$_6$ octahedral rotations and strain, respectively,
which we write as follows:
\begin{equation}
\label{eq:F_P}
\begin{aligned}
    F(P)=&A_P(P_x^2+P_y^2+P_z^2)+B_P(P_x^2+P_y^2+P_z^2)^2+\\
    & C_P(P_x^2P_y^2+P_y^2P_z^2+P_z^2P_x^2);
\end{aligned}
\end{equation}
\begin{equation}
\label{eq:F_R}
\begin{aligned}
    F(R)=&A_R(R_x^2+R_y^2+R_z^2)+B_R(R_x^2+R_y^2+R_z^2)^2+\\
    & C_R(R_x^2R_y^2+R_y^2R_z^2+R_z^2R_x^2);
\end{aligned}
\end{equation}
\begin{equation}
\label{eq:F_eta}
\begin{aligned}
    F(\eta)=& \frac{1}{2}C_{11}(\eta_{xx}^2+\eta_{yy}^2+\eta_{zz}^2)+C_{12}(\eta_{xx}\eta_{yy}+ \\ & \eta_{yy}\eta_{zz}+\eta_{zz}\eta_{xx})+ \frac{1}{2}C_{44}(\eta_{yz}^2+\eta_{xz}^2+\eta_{xy}^2).
\end{aligned}
\end{equation}
We truncate the expansion in $\mathbf{P}$ and $\mathbf{R}$ at the
fourth order, which is the minimum required to model structural
instabilities. In turn, we only consider harmonic terms for the
strains. 

Then, $F(P,R)$, $F(P,\eta)$ and $F(R,\eta)$ are the coupling terms,
which we write as:
\begin{equation}
\label{eq:F_PR}
    \begin{aligned}
    F(&P,R)=B_{PR}(P_x^2+P_y^2+P_z^2)(R_x^2+R_y^2+R_z^2)+ \\
    & C_{PR}(P_x^2R_x^2+P_y^2R_y^2+P_z^2R_z^2)+ \\
    & C'_{PR}(P_x P_y R_x R_y+P_y P_z R_y R_z+P_z P_x R_z R_x);
    \end{aligned}
\end{equation}
\begin{equation}
\label{eq:F_Peta}
    \begin{aligned}
    &F(P,\eta)= \gamma_{P111}(\eta_{xx}P_x^2+\eta_{yy}P_y^2+\eta_{zz}P_z^2) + \\ & \gamma_{P122}(\eta_{xx}(P_y^2+P_z^2)+ \eta_{yy}(P_z^2+P_x^2)+\eta_{zz}(P_x^2+P_y^2))+\\ & \gamma_{P423}(\eta_{yz}P_yP_z+\eta_{xz}P_zP_x+\eta_{xy}P_xP_y);
    \end{aligned}
\end{equation}
\begin{equation}
\label{eq:F_Reta}
    \begin{aligned}
    &F(R,\eta)= \gamma_{R111}(\eta_{xx}R_x^2+\eta_{yy}R_y^2+\eta_{zz}R_{z}^2) + \\ & \gamma_{R122}(\eta_{xx}(R_y^2+R_z^2)+ \eta_{yy}(R_z^2+R_x^2)+\eta_{zz}(R_x^2+R_y^2))+\\ & \gamma_{R423}(\eta_{yz}R_yR_z+\eta_{xz}R_zR_x+\eta_{xy}R_xR_y).
    \end{aligned}
\end{equation}
Here we restrict ourselves to the lowest-order symmetry-allowed
couplings between the considered order parameters. Note that, in these
equations, $A$, $B$, $C$, $C'$ and $\gamma$ are the material-dependent
expansion coefficients that we compute using DFT as detailed in
Sec.~\ref{subsec:parameters}. The coefficients $C_{11}$, $C_{12}$, and $C_{44}$ in Eq.~(\ref{eq:F_eta}), as well as $\gamma$ parameters in Eqs.~(\ref{eq:F_PR})-(\ref{eq:F_Reta}) are given in Voigt notation for compactness.
     

\subsection{Computing the potential parameters}
\label{subsec:parameters}

We now describe the approach to compute the expansion coefficients of
the Landau-like potential introduced in
Sec.~\ref{subsec:LD_potential}.  We mainly focus on an analytical
approach, but also discuss briefly a numerical scheme for comparison.

\begingroup
\renewcommand{\arraystretch}{1.3}
\begin{table*}
\caption{Polymorphs (labeled by $s$) included in the training set for computing the potential's coefficients and the notations for their polarizations $\mathbf{P}_s$, FeO$_6$ octahedral rotations $\mathbf{R}_s$ and components of the strain tensor $\bm{\eta}_s$. \label{tab:notations}}
\begin{tabular}{p{46pt}p{100pt}p{100pt}p{100pt}p{140pt}}
\hline
\hline
\multicolumn{2}{c}{\centering{Polymorph}} & \centering $\mathbf{P}$ & \centering  $\mathbf{R}$ & \centering $\bm{\eta}$
\tabularnewline
\hline
 \centering 1c & \centering P[001]c & \centering (0,0,$P_{1c}$) & \centering (0,0,0) & \centering (0,0,0,0,0,0)
\tabularnewline
 \centering 2c & \centering P[111]c & \centering ($P_{2c}$,$P_{2c}$,$P_{2c}$) & \centering (0,0,0) & \centering (0,0,0,0,0,0)
\tabularnewline
 \centering 3c & \centering R[001]c & \centering (0,0,0) & \centering (0,0,$R_{3c}$) & \centering (0,0,0,0,0,0)
\tabularnewline
 \centering 4c & \centering R[111]c & \centering (0,0,0) & \centering ($R_{4c}$,$R_{4c}$,$R_{4c}$) & \centering (0,0,0,0,0,0) 
\tabularnewline
 \centering 5c & \centering P[001]+R[001]c & \centering (0,0,$P_{5c}$) & \centering (0,0,$R_{5c}$) & \centering (0,0,0,0,0,0)
\tabularnewline
 \centering 6c & \centering P[111]+R[111]c & \centering ($P_{6c}$,$P_{6c}$,$P_{6c}$) & \centering ($R_{6c}$,$R_{6c}$,$R_{6c}$) & \centering  (0,0,0,0,0,0)
\tabularnewline
 \centering 7c & \centering P[11$\bar{1}$]+R[111]c & \centering ($P_{7c,\perp}$,$P_{7c,\perp}$,$P_{7c,\parallel}$) & \centering ($R_{7c,\perp}$,$R_{7c,\perp}$,$R_{7c,\parallel}$) & \centering (0,0,0,0,0,0)
\tabularnewline
 \centering 1 & \centering P[001] & \centering (0,0,$P_{1}$) & \centering (0,0,0) & \centering ($\eta_{1,\perp}$,$\eta_{1,\perp}$,$\eta_{1,\parallel}$,0,0,0)
\tabularnewline
 \centering 2 & \centering P[111] & \centering ($P_{2}$,$P_{2}$,$P_{2}$) & \centering (0,0,0) & \centering ($\eta_{2}$,$\eta_{2}$,$\eta_{2}$,$s_{2}$,$s_{2}$,$s_{2}$)
\tabularnewline
 \centering 3 & \centering R[001] & \centering (0,0,0) & \centering ($R_{3}$,0,0) & \centering ($\eta_{3,\parallel}$,$\eta_{3,\perp}$,$\eta_{3,\perp}$,0,0,0)
\tabularnewline
\centering 4 & \centering R[111] & \centering (0,0,0) &  \centering ($R_{4}$,$R_{4}$,$R_{4}$) & \centering ($\eta_{4}$,$\eta_{4}$,$\eta_{4}$,$s_{4}$,$s_{4}$,$s_{4}$)
\tabularnewline
 \centering 5 & \centering P[001]+R[001] & \centering (0,0,$P_5$) & \centering (0,0,$R_5$) & \centering ($\eta_{5,\perp}$,$\eta_{5,\perp}$,$\eta_{5,\parallel}$,0,0,0) 
\tabularnewline
 \centering 6 & \centering P[111]+R[111] & \centering ($P_{6}$,$P_{6}$,$P_{6}$) & \centering ($R_{6}$,$R_{6}$,$R_{6}$) & \centering ($\eta_{6}$,$\eta_{6}$,$\eta_{6}$,$s_{6}$,$s_{6}$,$s_{6}$)
\tabularnewline
\hline
\hline
\end{tabular}
\end{table*}


\subsubsection{Training set}
\label{subsubsec:training_set}

We first identify the states or polymorphs that we want our models to
describe. We consider the ground state as well as the low-energy polymorphs of the material, including the states that might be relevant for polarization switching. We thus define a training set of first-principles results
corresponding to the energies and structures of such polymorphs.

Before we continue, let us introduce a convenient notation for the
polymorphs we consider: we write ``P(R)[...]c'' or ``P[...]+R[...]c'',
where the first letter, P or R, indicates whether the structure
presents a polar distortion or FeO$_6$ octahedral tilts, respectively
(if both distortions appear, we indicate it by P+R); then, [001] or
[111] shows the axis along/about which the corresponding distortion is
oriented; finally, "c" indicates that the cubic cell is kept
fixed. Thus, for example, the polymorph P[001]c is characterized by a polar
distortion along the [001] direction and its cell is fixed to that of
cubic reference structure. For simplicity, we also introduce short
notations for all the polymorphs of interest, such as ``1c'' for the state
P[001]c. We summarize all the notations for the polymorphs and the
corresponding order parameters in Table \ref{tab:notations}.

\paragraph*{BiFeO$_3$} The starting point for constructing the training set is
the already-mentioned 40-atom supercell compatible with the G-type
antiferromagnetic order and the antiphase rotations of the FeO$_6$
octahedra. First, we run a DFT simulation to optimize the volume of the cubic phase of BiFeO$_{3}$ using this supercell. Next, we use the optimized structure to construct six polymorphs
(1c to 6c in Table~\ref{tab:notations}) by imposing the polar
distortion and/or antiphase octahedral rotation along/about either the
[001] or [111] directions while keeping the volume and shape of the
supercell fixed (the corresponding ionic displacement patterns are
illustrated in Fig.~\ref{fig:distortion_modes}). We use DFT to
optimize the ionic positions in these polymorphs and calculate the
energies $E_s$ of the resulting structures, where the index $s$ labels
polymorphs in the training set.  Additionally, we also consider the
state we call P[11$\bar{1}$]+R[111]c (7c); here, we impose a polar
distortion and octahedral tilts with amplitudes typical of BiFeO$_3$,
but oblique to each other. This structure does not correspond to a
special point of the energy landscape; therefore, we do not perform a
structural optimization and only compute its energy, which is
needed to obtain the coefficient $C_{PR}'$, as we will show in
Sec.~\ref{subsubsec:analytical}.

Next, we consider the first six polymorphs mentioned above (structures
1 to 6 in Table~\ref{tab:notations}), but now allowing for changes in
the shape and volume of the supercell (note we omit the ``c'' in the
notation).

In all cases we extract the displacements $\mathbf{u}_{Bi,s}$ ($u_{Bi,s}$ are in Angstrom) of the
Bi cations with respect to the corresponding O anion cages. We average
the values of these displacements over all Bi ions to obtain
$\mathbf{\bar{u}}_{Bi,s}$. Since the Bi off-centering largely
determines the electric polarization in BiFeO$_3$, we estimate
$\mathbf{P}_s$ for the considered polymorphs as
$\mathbf{P}_s=K_0\bar{\mathbf{u}}_{Bi,s}$, where
$K_0=P_0/\bar{u}_{Bi,6}$, $P_0$=0.58 C/m$^2$ and $\bar{\mathbf{u}}_{Bi,6}$ is the
average Bi off-centering in the ground state P[111]+R[111]. This
choice of $P_0$ ensures that the spontaneous polarization of the
P[111]+R[111] polymorph is $\mathbf{P_6}=P_0(1,1,1)$ which gives the
magnitude of $\mathbf{P_6}$ around its experimentally determined value
of 1~C/m$^2$.

Similarly, we compute the rotation angles of the FeO$_6$ octahedra
$\mathbf{R}_s$ about the pseudocubic axes, from which we obtain the
amplitude of the antiphase tilt pattern,
$\bar{\mathbf{R}}_s$. Finally, in the cases where the shape and
volume of the cell are allowed to relax, we extract also the
components of the strain tensor $\bm{\eta}_s$. The obtained results
constitute our training set, which is presented in Table~S1 of the
Supplementary Material.

\paragraph*{La-doped BiFeO$_3$} Experimentally, La dopants distribute
quasi-randomly in the BiFeO$_{3}$ lattice, so that the macroscopic
symmetry (cubic for the paraelectric phase, rhombohedral for the
ground state) is only recovered when a sufficiently large sample
volume is considered. Unfortunatately, reproducing such a situation in
a DFT calculation has a prohibitive computational cost; thus, here we
assume that a particular highly-ordered La arrangement, where the
dopants are as separated as possible from one another {\em and} which
respects the cubic symmetry of the reference lattice, is a good
approximation to the average experimental configuration. (For a 25~\%
La doping, the symmetric arrangement we use is shown in
Fig.~\ref{fig:crystal_structure}(b).) This approach allows us
to derive Landau potentials for doped materials, with the
experimentally relevant symmetry properties, from relatively
inexpesive DFT calculations. Admittedly, a careful (computationally
costly) validation of its accuracy remains for future work.

Having chosen a suitable, symmetric dopant arrangement, we optimize
the cubic cell of the reference paraelectric structure using
DFT. Next, we use this structure to construct the sets of polymorphs
1c to 7c and 1 to 6, in analogy to the case of pure BiFeO$_3$. For the
case of a 25~\% La composition, the obtained values of $\mathbf{P}_s$,
$\mathbf{R}_s$, $\bm{\eta}_s$ and $E_s$ of their optimized structures
are summarized in Table~S2 of the Supplementary Material.

Note that we encountered difficulties in constructing the training set
for La$_{1-x}$Bi$_x$FeO$_3$ compositions with an intermediate content
of La ($0<x<0.25$). For example, for La$_{0.125}$Bi$_{0.875}$FeO$_3$,
one can easily construct a paraelectric reference by subsituting a
single Bi atom in the supercell of Fig.~\ref{fig:crystal_structure}(a). However, we observed
that the P[001]+R[001]c and P[001]+R[001] polymorphs relax to the
lower symmetry phases displaying additional (and large) in-phase
rotations of the FeO$_6$ octahedra. These extra distortions are
secondary modes activated by the symmetry breaking associated to the
combination of polar and antiphase orders together with the considered
arrangement of La dopants. These distortions are not
expected to occur experimentally, as the La dopants are largely
disordered in real samples, and such in-phase tilts may occur locally
at most. Moreover, they cannot be treated within our simple potentials
(an explicit consideration of in-phase tilts would be required) and
complicate the definition of the training set. Hence, here we do not
compute models for such intermediate compositions. Nevertheless, as we
show in Sec.~\ref{subsec:intermediate}, suitable potentials can be
obtained by interpolation between those obtained for neighboring
(well-behaved) compositions.


\subsubsection{Analytical approach}
\label{subsubsec:analytical}

\begingroup \setlength{\tabcolsep}{9pt} 
\renewcommand{\arraystretch}{1.6} 
\begin{table}
\caption{Conditions used to derive the analytical expressions for the potential's coefficients. $E_s|\phi_{s,eq}$ denotes the energy of the polymorph $s$ corresponding to the equilibrium value of order parameter $\phi_{s}$. $\lambda=1$ m$^4$deg$^2$/C$^2$ is an {\sl ad hoc} coefficient used to balance the units for the terms in $f_{5c}$ and $f_{6c}$ (see text).}
\begin{tabular}{c c}
\hline
\hline
& \centering{Conditions} 
\tabularnewline
\hline
\centering{$A_P$} & \multirow{3}{*}{$\frac{\partial E_{1c}}{\partial P_{1c}}=0$; $\frac{\partial E_{2c}}{\partial P_{2c}}=0$; $E_{1c}|_{P_{1c,eq}}$; $E_{2c}|_{P_{2c,eq}}$}
\tabularnewline
\centering{$B_P$} & 
\tabularnewline
\centering{$C_P$} & 
\tabularnewline
\hline
\centering{$A_R$} & \multirow{3}{*}{$\frac{\partial E_{3c}}{\partial R_{3c}}=0$; $\frac{\partial E_{4c}}{\partial R_{4c}}=0$; $E_{3c}|_{R_{3c,eq}}$; $E_{4c}|_{R_{4c,eq}}$} 
\tabularnewline
\centering{$B_R$} & 
\tabularnewline
\centering{$C_R$} & 
\tabularnewline
\hline
\centering{$B_{PR}$} & \multirow{3}{*}{ $\frac{\partial E_{5c}}{\partial P_{5c}}+\lambda \frac{\partial E_{5c}}{\partial R_{5c}}=0$; $\frac{\partial E_{6c}}{\partial P_{6c}}+\lambda \frac{\partial E_{6c}}{\partial R_{6c}}=0$; $E_{7c}$ } 
\tabularnewline
\centering{$C_{PR}$} & 
\tabularnewline
\centering{$C'_{PR}$} & 
\tabularnewline
\hline
\centering{$\gamma_{P111}$} & $\frac{\partial E_{6}}{\partial \eta_{6}}=0$
\tabularnewline
\centering{$\gamma_{P122}$} & $\frac{\partial E_{1}}{\partial \eta_{1,\perp}}=0$; $\frac{\partial E_{2}}{\partial \eta_{2}}=0$ 
\tabularnewline
\centering{$\gamma_{P423}$} & $\frac{\partial E_{6}}{\partial s_{6}}=0$
\tabularnewline
\centering{$\gamma_{R111}$} & $\frac{\partial E_{3}}{\partial \eta_{3,\perp}}=0$
\tabularnewline
\centering{$\gamma_{R122}$} & $\frac{\partial E_{3}}{\partial \eta_{3,\perp}}=0$; $\frac{\partial E_{4}}{\partial \eta_{4}}=0$
\tabularnewline
\centering{$\gamma_{R423}$} & $\frac{\partial E_{4}}{\partial s_{4}}=0$
\tabularnewline
\hline
\hline
\end{tabular}
\label{tab:deriv_cond}
\end{table}

The approach introduced in this section allows full control of the
information used to compute the parameters of the potential.  To
achieve that, we derive an analytical expression for each of the
potential coefficients, in terms of $E_s$, $\mathbf{P}_s$,
$\mathbf{R}_s$ and $\bm{\eta}_s$ of the polymorphs in the training set
(see Sec.~\ref{subsubsec:training_set}). To obtain such formulas, we
use Eqs.~(\ref{eq:F_full}) - (\ref{eq:F_Reta}) and impose the
zero-derivative condition $\partial F/\partial \phi_{i,s}=0$, where
$\phi_{i,s}$ is the $i$th component of order parameter $\phi$
evaluated for polymorph $s$. Let us illustrate our procedure by
presenting in detail the case of the parameters $A_P$, $B_P$ and $C_P$ of
Eq.~\ref{eq:F_P}.

We consider two polar-only polymorphs with fixed cubic cell, P[001]c
(1c) and P[111]c (2c), and use the notation for their polarization
components introduced in Table \ref{tab:notations},
$\mathbf{P}_{1c}=(0,0,P_{1c})$ and
$\mathbf{P}_{2c}=(P_{2c},P_{2c},P_{2c})$.  Then, from
Eq.~(\ref{eq:F_P}) we can write the polymorph energies
\begin{equation}
\label{eq:E1c}
    E_{1c}=A_P P_{1c}^2+B_P P_{1c}^4
\end{equation}
and
\begin{equation}
\label{eq:E2c}
    E_{2c}=3A_P P_{2c}^2+3(3B_P+C_P) P_{2c}^4.
\end{equation}
By taking the derivatives of these energies with respect to $P_{1c}$
and $P_{2c}$, and setting them equal to zero, we obtain the following
equations for the equilibrium values of $P_{1c}$ and $P_{2c}$:
\begin{equation}
\label{eq:P1c}
P_{1c}^2=-\frac{A_P}{2B_P}
\end{equation}
and
\begin{equation}
\label{eq:P2c}
P_{2c}^2=-\frac{A_P}{2(3B_P+C_P)}.
\end{equation}
Then, by using these expressions in Eqs.~(\ref{eq:E1c}) and
(\ref{eq:E2c}), we obtain, respectively, $E_{1c}$ and $E_{2c}$ as 
functions of the parameters of the potential, such as
\begin{equation}
\label{eq:E1c_new}
E_{1c}=-\frac{A_P^2}{4B_P}
\end{equation}
and
\begin{equation}
\label{eq:E2c_new}
E_{2c}=-\frac{3A_P^2}{4(3B_P+C_P)}.   
\end{equation}
From Eq.~(\ref{eq:P2c}) one can see that $3B_P+C_P=-A_P/2P_{2c}^2$. By
using this in Eq.~(\ref{eq:E2c_new}), one can straightforwardly obtain
the analytical expression for $A_P$:
\begin{equation}
\label{eq:AP}
A_P=\frac{2E_{2c}}{3P_{2c}^2}.    
\end{equation}
From Eq. (\ref{eq:E1c_new}), in turn, one can obtain:
\begin{equation}
\label{eq:BP}
B_P=-\frac{A_P^2}{4E_{1c}}.    
\end{equation}
Finally, by combining Eqs. (\ref{eq:E2c_new}), (\ref{eq:AP}) and
(\ref{eq:BP}), we get:
\begin{equation}
\label{eq:CP}
C_P=\frac{3A_P^2}{4}\left(\frac{1}{E_{1c}}-\frac{1}{E_{2c}}\right).
\end{equation}
Since $E_{1c}$, $E_{2c}$, $P_{1c}$ and $P_{2c}$ are known from our
first-principles calculations described above, the coefficients $A_P$,
$B_P$ and $C_P$ can be directly computed using Eqs.~(\ref{eq:AP}),
(\ref{eq:BP}) and (\ref{eq:CP}), respectively.

Similarly, we can derive the analytical expressions for the remaining
coefficients of our potential. The specific conditions and properties
used in the derivation are summarized in Table \ref{tab:deriv_cond},
and the resulting expressions are:
\begin{equation}
  A_R=\frac{2E_{4c}}{3R^2_{4c}};  
\end{equation}
\begin{equation}
  B_R=\frac{-A_R^2}{4E_{3c}};  
\end{equation}
\begin{equation}
  C_R=\frac{3A_R^2}{4}\left(\frac{1}{E_{3c}}-\frac{1}{E_{4c}}\right);
\end{equation}
\begin{equation}
B_{PR}=\frac{f_{7c}-f_{5c}(C_{7c}-C'_{7c})-\frac{1}{3}C'_{7c}f_{6c}}{B_{7c}-C_{7c}-2C'_{7c}}, 
\end{equation}
where
\begin{equation}
    B_{7c}=(2P_{7c,\perp}^2+P_{7c,\parallel}^2)(2R_{7c,\perp}^2+R_{7c,\parallel}^2),
\end{equation}
\begin{equation}
    C_{7c}=2P_{7c,\perp}^2R_{7c,\perp}^2+P_{7c,\parallel}^2R_{7c,\parallel}^2,
\end{equation}
\begin{equation}
\begin{aligned}
    C'_{7c}=&P_{7c,\perp}^2R_{7c,\perp}^2+\\ & 2P_{7c,\perp}P_{7c,\parallel}R_{7c,\perp}R_{7c,\parallel},
\end{aligned}
\end{equation}
\begin{equation}
    f_{5c}=-\frac{A_P+\lambda A_R+2B_P P_{5c}^2+2\lambda B_R R_{5c}^2}{\lambda P_{5c}^2+R_{5c}^2}
\end{equation}
and
\begin{equation}
\begin{aligned}
    f_{6c}=&-\frac{1}{\lambda P_{6c}^2+R_{6c}^2}(3A_P+3\lambda A_R+
    \\ & 6(3B_P+C_P)P_{6c}^2+6\lambda (3B_R+C_R)R_{6c}^2),
\end{aligned}
\end{equation}
where $\lambda=1$~m$^4$deg$^2$/C$^2$ is an {\sl ad hoc} coefficient
that allows us to combine two zero-derivative conditions (for
polarization and tilts, respectively) into only one. Further, we have
\begin{equation}
\begin{aligned}
    f_{7c}=& E_{7c}-(A_P(2P_{7c,\perp}^2+P_{7c,\parallel}^2)+ \\ &
    B_P(2P_{7c,\perp}^2+P_{7c,\parallel}^2)^2+ \\ &
    C_P(P_{7c,\perp}^4+2P_{7c,\perp}^2P_{7c,\parallel}^2)+
    \\ &A_R(2R_{7c,\perp}^2+R_{7c,\parallel}^2)+ \\ &
    B_R(2R_{7c,\perp}^2+R_{7c,\parallel}^2)^2+ \\ &
    C_R(R_{7c,\perp}^4+2R_{7c,\perp}^2R_{7c,\parallel}^2));
\end{aligned}
\end{equation}
\begin{equation}
    C_{PR}=f_{5c}-B_{PR};
\end{equation}
\begin{equation}
C'_{PR}=\frac{1}{C'_{7c}}\left(f_{7c}-B_{7c}B_{PR}-C_{7c}C_{PR}\right);    
\end{equation}
\begin{equation}
    \gamma_{P122}=\frac{(C_{12}-C_{11})\eta_{1,\perp}}{P_1^2} -\frac{C_{12}\eta_{2}}{P_{2}^2};
\end{equation}
\begin{equation}
\begin{aligned}
    \gamma_{P111}= &-2\gamma_{P122}- \\ & \left(\eta_{6}+\frac{(\gamma_{R111}+2\gamma_{R122})R_{6}^2}{C_{11}+2C_{12}}\right)
    \frac{C_{11}+2C_{12}}{P_6^2};
\end{aligned}
\end{equation}
\begin{equation}
    \gamma_{P423}=-\left(s_6+\frac{\gamma_{R423}R_6^2}{C_{44}}\right)\frac{C_{44}}{P_6^2};
\end{equation}
\begin{equation}
    \gamma_{R111}=\frac{2(C_{11}-C_{12})\eta_{3,\perp}}{R_{3}^2}-\frac{C_{11}\eta_{4}}{R_{4}^2};
\end{equation}
\begin{equation}
    \gamma_{R122}=\frac{(C_{12}-C_{11})\eta_{3,\perp}}{R_{3}^2}-\frac{C_{12}\eta_{4}}{R_{4}^2};
\end{equation}
and
\begin{equation}
    \gamma_{R423}=-\frac{C_{44}s_{4}}{R_{4}^2}.
\end{equation}
Note, that it is possible to choose other conditions, different from
those in Table~\ref{tab:deriv_cond}, to derive the expressions for the
model parameters. For example, $\gamma_{P111}$ might be obtained from
the energies and structures of polar-only polyrmorphs, in analogy to
what we do for $\gamma_{R111}$ usign tilt-only polymorphs. However, we
find that this choice yields shear strains with incorrect sign for the
P[111]+R[111] ground state of BiFeO$_3$. By contrast, the condition we
use to compute $\gamma_{P111}$ (i.e., $\partial E_6/\partial
\eta_6=0$) includes the information about the ground state and
corrects this problem. These difficulties reflect the simplicity of
our low-order polynomial model, which can account (exactly) for only a
small number of properties.

Finally, the elastic constants $C_{11}$, $C_{12}$ and $C_{44}$ are
calculated directly from DFT.

\subsubsection{Numerical approach}
\label{subsubsec:least_squares}

The numerical approach that we introduce in this section allows to
compute the potential coefficients using the information from all
considered structural polymorphs. We focus here on the case of pure
BiFeO$_3$, noting that exactly the same procedure can be applied to
La$_{0.25}$Bi$_{0.75}$FeO$_3$.

We work with the BiFeO$_3$ polymorphs from the training set introduced
in Sec.~\ref{subsubsec:training_set}, namely, 1c to 7c and 1 to 6 of
Table~\ref{tab:notations}. Based on the energies ($E_s$) and
structural parameters ($\mathbf{P}_s$, $\mathbf{R}_s$, and
$\bm{\eta}_s$) obtained from DFT, we construct an overdetermined
system of linear equations, with the potential parameters as unknowns,
using the expressions for the energy and zero derivatives
corresponding to all polymorphs. (For the 7c state, the
zero-derivative condition does not apply. Also, we use the elastic
constants $C_{11}$, $C_{12}$ and $C_{44}$ directly obtained from DFT.)

We find that the potential obtained using this approach provides
less accurate predictions for the properties of the
low-energy polymorphs of BiFeO$_3$ as compared to the analytical
approach introduced in Sec.~\ref{subsubsec:analytical} (see details in
Sec. SII of the Supplementary material). More specifically, the
numerically-determined potential does a better job at reproducing some
features (e.g., the polarization of the P[001] state) that we
disregard in our analytical approach. In turn, it is less accurate
when it comes to capture some critical properties (e.g., the ground
state energy). We conclude that, while this fitting approach might
work well for more complete, higher-order potentials (for example,
such as the one introduced in Ref.~\onlinecite{marton2017}), it seems
less suitable for computing the parameters of our low-order
model. Therefore, in the following, we are going to discuss only the
results obtained using the analytical approach.


\section{Results}
\label{sec:results}
\subsection{BiFeO$_3$}
\label{subsec:results_BFO}

In this section, we analyze how accurately the potential introduced in
Sec.~\ref{subsec:LD_potential} predicts the properties of BiFeO$_3$
polymorphs.  We begin by computing the coefficients of the potential
following the analytical approach described in
Sec.~\ref{subsubsec:analytical}. The resulting values are presented in
Table~\ref{tab:LD_parameters}. Next, we use the computed potential to
calculate the equilibrium properties ($\mathbf{P}_s$, $\mathbf{R}_s$,
$\bm{\eta}_s$, and $E_s$) of the polymorphs 1c to 6c and 1 to 6.
Since in these polymorphs the form of $\mathbf{P_s}$ is either
$(0,0,P_s)$ or $(P_s,P_s,P_s)$ (the same holds for $\mathbf{R}_s$), in
the following we will discuss single components of $\mathbf{P}_s$ and
$\mathbf{R}_s$ ($P_s$ and $R_s$, respectively).  We plot the values of
$P_s$, $R_s$ and $E_s$ predicted using our potential versus their DFT
counterparts as shown in Fig. \ref{fig:PRE_BFO} (all these values, as
well as the components of $\bm{\eta}_s$ are also presented in Table S1
of the Supplementary Material). We note that, if the model prediction
and DFT value match exactly, the corresponding point lays on the black
dashed line.

\begingroup
\setlength{\tabcolsep}{9pt} 
\renewcommand{\arraystretch}{1.2} 
\begin{table}
\caption{Coefficients of our Landau-like potentials calculated for BiFeO$_3$ (BFO) and
  La$_{0.25}$Fe$_{0.75}$FeO$_3$ (LBFO) using the analytical approach
  introduced in Sec. \ref{subsubsec:analytical}. We give energies per 5-atom perovskite unit cell.}
\resizebox{\columnwidth}{!}{
\begin{tabular}{c S[table-format=-1.5] S[table-format=-1.5] c}
\hline
\hline
& \centering{BFO} & \centering{LBFO} & Units
\tabularnewline
\hline
\centering{$A_P$} & -1.747 & -1.674 & \centering{$\times10^{-19}$, J m$^4$ C$^{-2}$}
\tabularnewline
\centering{$B_P$} & 1.070 & 1.286 & \centering{$\times10^{-19}$, J m$^8$ C$^{-4}$}
\tabularnewline
\centering{$C_P$} & -7.486 & -6.212 & \centering{$\times10^{-20}$, J m$^8$ C$^{-4}$}
\tabularnewline
\centering{$A_R$} & -8.555 & -7.560 & \centering{$\times10^{-22}$, J  deg$^{-2}$}
\tabularnewline
\centering{$B_R$} & 2.169 & 1.962 & \centering{$\times10^{-24}$, J  deg$^{-4}$}
\tabularnewline
\centering{$C_R$} & -1.240 & -0.848 & \centering{$\times10^{-24}$, J deg$^{-4}$}
\tabularnewline
\centering{$C_{11}$} & 1.833 & 1.754 & \centering{$\times10^{-17}$, J}
\tabularnewline
\centering{$C_{12}$} & 7.301 & 11.280 &  \centering{$\times10^{-18}$, J}
\tabularnewline
\centering{$C_{44}$} & 4.600 & 4.262 & \centering{$\times10^{-18}$, J}
\tabularnewline
\centering{$B_{PR}$} &  1.121 & 1.183 & \centering{$\times10^{-21}$, J m$^4$ C$^{-2}$ deg$^{-2}$}
\tabularnewline
\centering{$C_{PR}$} & -3.437 & -3.319  & \centering{$\times10^{-22}$, J m$^4$ C$^{-2}$ deg$^{-2}$}
\tabularnewline
\centering{$C'_{PR}$} & -2.245 & -2.219 & \centering{$\times10^{-21}$, J m$^4$ C$^{-2}$ deg$^{-2}$}
\tabularnewline
\centering{$\gamma_{P111}$} & -9.444 & -7.866 & \centering{$\times10^{-19}$, J m$^4$ C$^{-2}$}
\tabularnewline
\centering{$\gamma_{P122}$} & -1.557 & -4.898 & \centering{$\times10^{-19}$, J m$^4$ C$^{-2}$}
\tabularnewline
\centering{$\gamma_{P423}$} & -3.232 & -3.359 & \centering{$\times10^{-19}$, J m$^4$ C$^{-2}$}
\tabularnewline
\centering{$\gamma_{R111}$} & -1.178 & 0.482 & \centering{$\times10^{-21}$, J  deg$^{-2}$}
\tabularnewline
\centering{$\gamma_{R122}$} & 1.158 & -10.026 & \centering{$\times10^{-22}$, J   deg$^{-2}$}
\tabularnewline
\centering{$\gamma_{R423}$} & 1.155 & 1.022 & \centering{$\times10^{-21}$, J   deg$^{-2}$}
\tabularnewline
\hline
\hline
\end{tabular}
}
\label{tab:LD_parameters}
\end{table}

First, we discuss the BiFeO$_3$ polymorphs with the fixed cubic cell
(no strain relaxation). From Figs.~\ref{fig:PRE_BFO}(a) and
\ref{fig:PRE_BFO}(b) one can see that, for these polymorphs, our model
predicts $P_s$ and $R_s$ in nearly perfect agreement with DFT. As
shown in Fig.~\ref{fig:PRE_BFO}(c), it also reproduces accurately
their energies and, therefore, their relative stability. Indeed, among
the structures having only polar distortion (P[001]c and P[111]c), the
one with $\mathbf{P} \parallel [111]$  is lower in energy
according to both model and DFT. The same holds for the structures
having only FeO$_6$ octahedral rotations (R[111]c is lower in energy
than R[001]c). Overall, the lowest energy structure is P[111]+R[111]c,
in which both distortions coexist and oriented along/about [111]. Here, one should keep in mind that DFT information about
these polymorphs is explicitly used to compute the model parameters
(see Table~\ref{tab:deriv_cond}), hence the agreement is not
surprising. Nevertheless, the potential does provide accurate
predictions for quantities that are not considered in its derivation
(e.g., $P$ of P[001]c, $R$ of R[001]c, and $E_s$ of P[001]+R[001]c and
P[111]+R[111]c).

\begin{figure*}
    \centering
    \includegraphics[width=0.99\linewidth,trim=0cm 0cm 0cm 0cm]{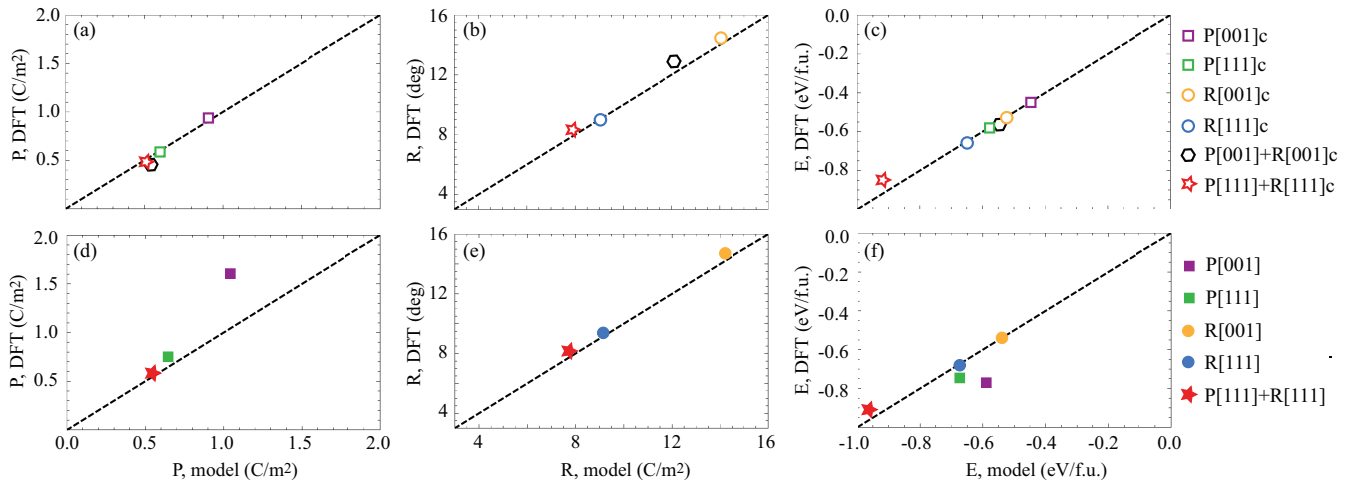}
    \caption{Structural properties and energies of BiFeO$_3$
      polymorphs predicted using the Landau-like potential and plotted
      versus their corresponding DFT values. The top row shows the results obtained for
      the polymorphs with fixed cubic supercell (no
      strain), while the bottom row shows the properties of the polymorphs with
      fully relaxed cells. Panels (a)
      and (d) show the electric polarization $P$, (b) and (e)  the
      FeO$_6$ octahedral rotations $R$, (c) and (f) the energies
      $E$. The polymorph P[001]+R[001] is not
      shown in panels (d) - (f), since its fully
      relaxed structure has very small FeO$_6$ octahedral rotations
      ($R=0.066^o$) and our model predicts it to be zero, therefore reducing to the state P[001].}
    \label{fig:PRE_BFO}
\end{figure*}

Next, we consider the BiFeO$_3$ polymorphs with allowed strain
relaxation (Figs.~\ref{fig:PRE_BFO}(d)-(f)). In this case, our
potential also provides accurate predictions for all considered
quantities for the most of the considered polymorphs; in particular,
it yields the correct ground state of BiFeO$_3$ (P[111]+R[111]). There
is only one polyrmorph for which the model is less accurate, namely,
P[001]. In this case, the DFT optimized structure has a large
distortion along the $z$ axis (the $c/a$ ratio is approximately 1.27),
accompanied by a large $P_z$; this is usually called supertetragonal
phase \cite{bea2009,zeches2009}. This behavior is not well captured by our
potential, as it underestimates the polarization and strains
components compared to the DFT values ($P_z=1.039$ versus
1.624~C/m$^2$; $\eta_{xx}=-0.012$ versus $-0.044$; $\eta_{zz}=0.065$
versus 0.216). This issue is also reflected in the energy predicted for
this phase. From the DFT results one can see that, among the
polymorphs with only polar distortion, the strain relaxation
stabilizes the supertetragonal phase over the rhombohedral one (P[001]
is lower than P[111] by 0.023~eV/f.u).  Our model does predict the
energy lowering of P[001] state due to the strain relaxation (the
negative $\gamma_{P111}$ coupling results in large $P_z$ and
$\eta_{zz}$). However, since it underestimates $P_z$ and $\eta_{zz}$,
this energy reduction is not enough to stabilize P[001]
over P[111]. Note that these deficiencies were to be expected, as we
decided to use a minimal amount of DFT information on the
supertetragonal phase when deriving the parameters of our model (see
Table~\ref{tab:deriv_cond}), because this state is not relevant for
our ultimate purpose of studying polarization switching in the rhombohedral phase of
BiFeO$_3$. Moreover, we checked that, if we try to capture the
supertetragonal $c/a$, this makes it difficult to obtain a correct
prediction for the ground state, as the P[001] state tends to become
dominant.


\subsection{La$_{0.25}$Bi$_{0.75}$FeO$_3$}
\label{subsec:results_LBFO} 

Now we discuss the case of La$_{0.25}$Bi$_{0.75}$FeO$_3$. We first
compute the parameters of the potential using the analytical
expressions in Sec.~\ref{subsubsec:analytical}. The resulting values
are presented in Table~\ref{tab:LD_parameters}. Next, we compare the
model predictions and DFT values for our considered polymorphs in
Fig.~\ref{fig:PRE_LBFO} (these results are also summarized in Table~S2
of the Supplementary Material, together with the corresponding
strains).

\begin{figure*}
    \centering
    \includegraphics[width=0.99\linewidth,trim=0cm 0cm 0cm 0cm]{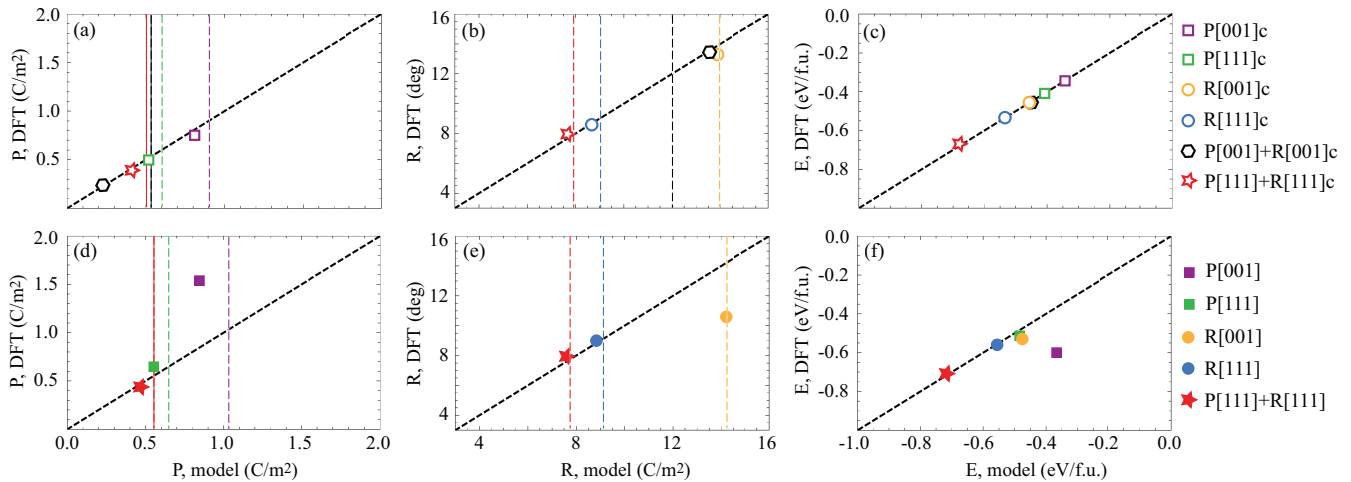}
    \caption{Structural properties and energies of La$_{0.25}$Bi$_{0.75}$FeO$_3$
      polymorphs predicted using the Landau-like potential and plotted
      versus their corresponding DFT values. The top row shows the results obtained for
      the polymorphs with fixed cubic supercell (no
      strain), while the bottom row shows the properties of the polymorphs with
      fully relaxed cells. Panels (a)
      and (d) show the electric polarization $P$, (b) and (e)  the
      FeO$_6$ octahedral rotations $R$, (c) and (f) the energies
      $E$. The polymorph P[001]+R[001] is not
      shown in panels (d) - (f), since its fully
      relaxed structure has very small FeO$_6$ octahedral rotations
      ($R=0.006^o$) and our model predicts it to be zero, therefore reducing to the state P[001]. The vertical dashed lines in panels (a), (b), (d) and (e) indicate the corresponding quantities for pure BiFeO$_3$.}
    \label{fig:PRE_LBFO}
\end{figure*}

Let us first consider the polymorphs with fixed cubic cell
(Fig.~\ref{fig:PRE_LBFO}(a)-(c)). Our potential provides very accurate
predictions for polarizations and tilts, similarly to the case of pure
BiFeO$_3$. The energy and relative stability of these polymorphs is
also well captured by the model. Indeed, for polar-only structures
both the model and DFT predict the rhombohedral P[111]c state to be
lower in energy than the tetragonal P[001]c phase. The same holds for
the polymorphs having only FeO$_6$ rotations: R[111]c is lower in
energy than R[001]c. Note that the energy difference between the
structures with tetragonal and rhombohedral phases are reduced
compared to the case of pure BiFeO$_3$. The lowest-energy phase is
P[111]+R[111]c, where polarization and tilts coexist.

For the polymorphs in which shape and volume of the cell are allowed
to relax, we observe the following. First, the model predicts accurate
values of the polarization in all cases except for P[001]. Indeed, for
the supertetragonal state, the predicted $P_s$ and $\eta_{zz}$ are
underestimated relative to the DFT values. This issue is also
reflected in the energy of the polymorph: our potential predicts
P[001] to be the highest-energy state, while DFT shows that this phase
is the second-lowest in energy, right above the P[111]+R[111] ground
state.  Additionally, we find that the tilts are accurately predicted
by our model for all polymorphs except R[001]; in that case, the tilt
amplitude and the strains are  exaggerated compared to the DFT
values. Note that, as it was the case for pure BiFeO$_{3}$, these
deficiencies are the result of the limited amount of DFT information
on states P[001] and R[001] that was used to derive the parameters of
our model.

\subsection{Intermediate compositions}
\label{subsec:intermediate}

In this section we demonstrate how our potentials can be used to study
La$_{1-x}$Bi$_x$FeO$_3$ with intermediate La content, $0<x<0.25$. We
focus on the case of La$_{0.125}$Bi$_{0.875}$FeO$_3$ and check whether
the properties of the polymorphs from the training set can be
predicted by linear interpolation between BiFeO$_3$ and
La$_{0.25}$Bi$_{0.75}$FeO$_3$.

We consider two types of interpolation. First, we construct a model
for $x=0.125$ with coefficients obtained from interpolation of the
corresponding values for the $x=0$ and $x=0.25$ cases. Using this
model, we can easily predict the properties ($P_s$, $R_s$ and $E_s$)
of all the polymorhps in the training set. Second, we derive the very
same properties by direct interpolation of the values obtained at
$x=0$ and $x=0.25$. In Fig.~\ref{fig:PRE_LBFO_0125} we compare the
quantities thus obtained, and also include the corresponding DFT
values for the polymorphs for which the information is available (see
figure caption). We find that both interpolation approaches yield very
similar preditions. Further, the agreement with DFT is good except for
the supertetragonal P[001] phase, where our predictions suffer from
the issues discussed above. Hence, we conclude that our models give us
a way to treat compounds with intermediate compositions.

\section{Discussion}
\label{sec:discussion}
Let us now discuss the physical insights that our models provide.

\subsection{P-R coupling}
\label{subsec:PRcoupling}

\begin{figure*}
    \centering
    \includegraphics[width=0.99\linewidth,trim=0cm 0cm 0cm 0cm]{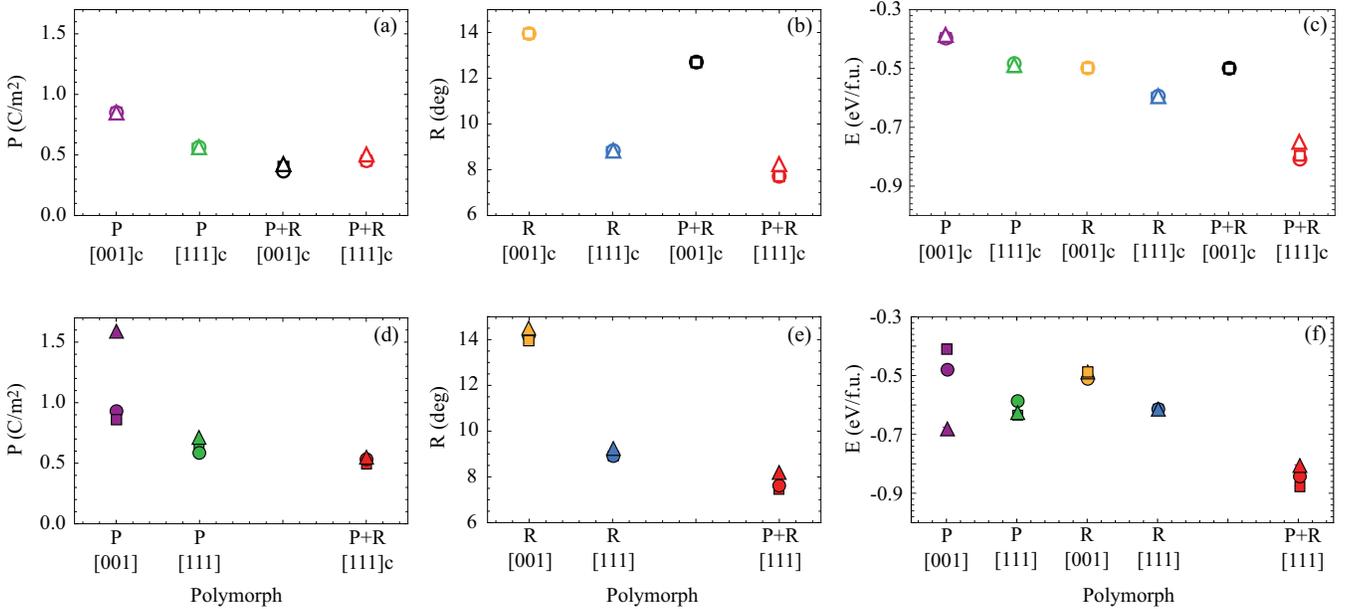}
    \caption{Structural properties and energies of La$_{0.125}$Bi$_{0.875}$FeO$_3$ polymorphs. Squares show the model predictions for $P$, $R$ and $E$ obtained using the  interpolated  potential's coefficients, circles - the values obtained by direct interpolation of the corresponding results between pure BiFeO$_3$ and La$_{0.25}$Bi$_{0.75}$FeO$_3$, triangles - DFT values. The top row shows the results for fixed cubic cell (no strain), the bottom row - the results after a full strain relaxation.  Panels (a) and (d) show the electric polarization $P$, (b) and (e) the FeO$_6$ octahedral rotations $R$, (c) and (f) the energies $E$. A missing DFT data point in the plot means that the  structure of the corresponding polymorph developed non-negligible additional distortions during the DFT optimization, as described in Sec. \ref{subsubsec:training_set}.}
    \label{fig:PRE_LBFO_0125}
\end{figure*}

As it is well known from both experiments and computations, and
correctly captured by our models, the ground state of BiFeO$_3$ has
rhombohedral symmetry with $\mathbf{P}\parallel[111]$ and
$\mathbf{R}\parallel[111]$. It is interesting to note, though, that
the DFT energy of the polar-only BiFeO$_3$ polymorph P[001] is lower
than that of P[111]. By contrast, among the polymorphs having only
FeO$_6$ octahedral tilts, R[111] is the lowest-energy structure. These
observations yield one important conclusion: that the rhombohedral
symmetry of the BiFeO$_{3}$'s ground state critically depends on the
presence of the octahedral tilts, as in their absence the material
would be tetragonal.

In order to understand how the rhombohedral ground state of BiFeO$_3$
comes about, we consider the $F(P,R)$ part of our potential
(Eq.~\ref{eq:F_PR}) describing the coupling between polarization and
octahedral rotations. The second term in Eq.~(\ref{eq:F_PR}), with
$C_{PR}<0$ (see Table \ref{tab:LD_parameters}), favors states where
$\mathbf{P}$ and $\mathbf{R}$ are along/about any $\langle111\rangle$
direction, as for example, $\mathbf{P} \parallel [111]$ and
$\mathbf{R} \parallel [\bar{1}1\bar{1}]$. In turn, the third term,
with $C'_{PR}<0$, favors phases where the FeO$_{6}$ tilts are about
the axis defined by the polarization. Overall, these couplings lead
$\mathbf{P}$ and $\mathbf{R}$ to appear together and aligned
along/about the same $\langle111\rangle$ direction. Thus, these are
the interactions driving the stabilization of the ground state phase
of BiFeO$_{3}$, rhombohedral and with co-existing polarization and tilts.

Does this mean, however, that $\mathbf{P}$ and $\mathbf{R}$ cooperate
in BiFeO$_3$? We address this question by considering the energy
diagram presented in Fig.~\ref{fig:en_diag}. Here we show the energies of
the P[111]c, R[111]c, and P[111]+R[111]c polymorphs as given by our
model ($E_{2c}$, $E_{4c}$ and $E_{6c}$, respectively, see Table
\ref{tab:energies_BFO}). We also show the energy of a virtual state in
which $\mathbf{P}\parallel [111]$ coexists with
$\mathbf{R}\parallel[111]$ {\em but} where these order parameters are not
coupled. The energy of this virtual state is simply given by
$E_{2c}+E_{4c}$, taking the cubic phase as the zero of
energy. Clearly, the P[111]+R[111]c polymorph is higher in energy than
the non-interacting virtual state and their energy difference arises
from the coupling between $\mathbf{P}$ and $\mathbf{R}$ in
P[111]+R[111]c structure. To understand this, we again consider the
term $F(P,R)$ (Eq.~\ref{eq:F_PR}) with the corresponding coefficients
$B_{PR}$, $C_{PR}$ and $C'_{PR}$ presented in Table
\ref{tab:LD_parameters} for BiFeO$_3$. For the polymorph
P[111]+R[111]c, we have $P_x=P_y=P_z=P_{6c}$ and $R_x=R_y=R_z=R_{6c}$;
therefore, $F_{6c}(P,R)= 3(3B_{PR}+C_{PR}+C'_{PR})P_{6c}^2R_{6c}^2$ for this
state. In this expression, $3B_{PR}>0$ dominates over
$C_{PR}+C'_{PR}<0$ and leads to a ground state energy that is higher than that of the virtual non-interacting state.

Thus, we find that, overall, the $\mathbf{P}$ and $\mathbf{R}$ order
parameters compete in BiFeO$_{3}$ ($B_{PR}>0$
dominates). Nevertheless, the polar and tilt instabilities are
so strong that this repulsive interaction is not enough to prevent
them from occurring simultaneously. Further, the
$\mathbf{P}$-$\mathbf{R}$ competition is minimized when the order
parameters are oriented along/about the same $\langle 111 \rangle$ axis
($C_{PR},C'_{PR}<0$), which yields the rhombohedral ground state phase
of BiFeO$_3$.

\subsection{Effects of La doping}

Let us first consider how La doping affects the electric
polarization. From Figs.~\ref{fig:PRE_LBFO}(a) and
\ref{fig:PRE_LBFO}(d) one can see that, for all considered polar
polymorphs in the training set, a 25\% La doping leads to reduction of
$P$. Indeed, for the polymorphs with fixed cubic cell
(Fig.~\ref{fig:PRE_LBFO}(a)) we obtain a reduction of $P$ by $11-19\%$
for P[001]c, P[111]c and P[111]+R[111]c, and an even larger reduction
for P[001]+R[001]c ($\approx59\%$). When we allow the cell to relax (Fig.~\ref{fig:PRE_LBFO}(d)),
the obtained $P$ reduction is in the range of $5-20\%$.

Next, let us turn to the effect of La doping on the FeO$_6$ octahedral
tilts. As one can see from Figs.~\ref{fig:PRE_LBFO}(b) and
\ref{fig:PRE_LBFO}(e), the presence of 25\% La has a relatively small
effect (reduction) in the amplitude of tilts. More precisely, we find
that the R[001]c, R[111]c and P[111]+R[111]c polymorphs with fixed
cubic cell present $1-4\%$ smaller $R$ compared to pure BiFeO$_3$. The
exception is the P[001]+R[001]c state, where La doping leads to an
increase in $R$ by 12\%. Finally, when we allow the cell to relax, we
find a $0.06-4\%$ reduction of $R$ in all considered polymorphs. 

\begin{table}
\caption{Energies $E_s$ of BiFeO$_3$ polymorphs calculated using DFT and predicted by the Landau-like potential introduced in this work (the coefficents of the potential are obtained using the analytical approach described in Sec.~\ref{subsubsec:analytical}). Energy values are relative to the energy of the reference cubic phase and given in eV per 5-atom unit cell. \label{tab:energies_BFO}}
\begin{tabular}{cccc}
\hline
\hline
 \multicolumn{2}{c}{Polymorph} & $E_s$ (DFT) & $E_s$ (Model) 
 \tabularnewline
\hline
 1c & P[001]c & -0.445 & -0.445
\tabularnewline
 2c & P[111]c & -0.580 & -0.580
\tabularnewline
 3c & R[001]c & -0.527 & -0.527
\tabularnewline
 4c & R[111]c & -0.651 & -0.650
\tabularnewline
 5c & P[001]+R[001]c & -0.556 & -0.546
\tabularnewline
 6c & P[111]+R[111]c &  -0.853 & -0.919
\tabularnewline
 1 & P[001] &  -0.764 & -0.589
\tabularnewline
 2 & P[111] &  -0.741 & -0.677
\tabularnewline
 3 & R[001] &  -0.536 & -0.540
\tabularnewline
 4 & R[111] &  -0.679 &  -0.671
\tabularnewline
 5 & P[001]+R[001] & -0.764 & -0.593
\tabularnewline
 6 & P[111]+R[111] & -0.909 & -0.967
\tabularnewline
\hline
\hline
\end{tabular}
\end{table}

\begin{figure}
    \centering
    \includegraphics[width=0.95\linewidth,trim=0cm 0cm 0cm 0cm]{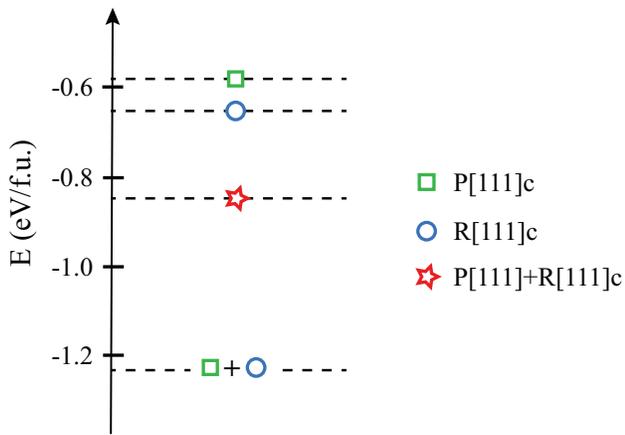}
    \caption{Model energies of several BiFeO$_3$ polymorphs. The lowest energy configuration corresponds to the virtual state in which $\mathbf{P}\parallel [111]$ and $\mathbf{R}\parallel [111]$ coexist but are not coupled (see text).}
    \label{fig:en_diag}
\end{figure}

Our models allow us to rationalize the most important results
described above. Let us start by noting that the
$\mathbf{P}$-$\mathbf{R}$ couplings ($B_{PR}$, $C_{PR}$ and $C_{PR}'$
in $F(P,R)$) are not significantly affected by the doping. Hence, they
do not play a significant role to explain the La-induced effects. 

Indeed, the effects of La doping on the polarization are essentially
captured by the changes in the $F(P)$ term of the potential
(Eq.~\ref{eq:F_P}). As shown in Table~\ref{tab:LD_parameters}, we find
that $A_P$ (quadratic coupling) is reduced in magnitude upon doping,
indicating a weaker ferroelectric instability of the cubic
phase. Additionally, both $B_P$ and $C_P$ increase and the relevant
combination, $3B_{P}+C_{P}>0$, becomes larger;
hence, the quartic couplings have a stronger effect on the energy
landscape compared to pure BiFeO$_3$. All these changes cooperate to
yield shallower ferroelectric energy wells associated to $F(P)$ for
La-doped BiFeO$_3$, with smaller equlibrium polarization and lower
energy barrier between states of opposite $\mathbf{P}$. Note that this
is consistent with previous studies on the effect of La-doping on the
switching characteristics of BiFeO$_{3}$ \cite{prasad2020,parsonnet2020}.

As regards the tilt energy given by $F(R)$,
Table~\ref{tab:LD_parameters} shows that the presence of La weakens
the cubic-phase instability ($A_{R}$ becomes less negative); by
contrast, the quartic term ($3B_{R}+C_{R}>0$) gets reduced upon
doping, thus favoring larger tilts. These changes oppose each other,
and result in the generally observed moderate reduction in the
amplitude of the FeO$_{6}$ rotations.

Finally, as shown in Fig.~\ref{fig:PRE_LBFO}, the P[001]+R[001]c case
is peculiar, as it presents the largest reduction in $P$ (about 59~\%)
and is the only one displaying an increase of $R$ (about 12~\%). We
can rationalize this result by noting that, for this state, the
quartic part of the energy in $F(P)$ (resp. $F(R)$) is controlled by
the $B_{P}$ (resp. $B_{R}$) coupling alone. Upon doping $B_{P}$ grows
($B_{R}$ decreases), which favors smaller polarizations (larger
tilts). Further, because of the strong competition between
polarization and tilts in tetragonal states ($B_{PR}>0$; $C_{PR}$ and $C'_{PR}$ do not contribute), the changes
get particularly large in the case of P[001]+R[001]c. Note also a
subtle difference between P[001]+R[001]c and P[111]+R[111]c. In the
latter case, the relevant quartic parameter for the tilts is
$3B_{R}+C_{R}$, and the La-induced decrease in $B_{R}$ is partly
compensated by the increase in $C_{R}$; as a result, the tilts do not
grow at all (recall $A_{R}<0$ grows upon doping) and the decrease of
the polarization is relatively small.

Note that all these observations are consistent with what we know
about the atomistic origin of the polar and tilt instabilities in
BiFeO$_{3}$. The former rely on the presence of stereochemically
active $6s$ lone pairs in the Bi$^{3+}$ cations; hence, their partial
substitution by lone-pair-free La cations naturally leads to smaller
polarizations. The latter are mainly controlled by the ionic radius of
the Bi$^{3+}$ cation; since La$^{3+}$ is similar in size, the doping
leaves $R$ largely unaffected.

\section{Conclusions}
\label{sec:conclusions}

In summary, we have introduced the simplest, lowest-order Landau-like
potential for BiFeO$_3$ and related compounds, as well as methods that
allow to compute the potential parameters from Density Functional
Theory (DFT). More precisely, we have derived analytical expressions
for all the model coefficients as functions of the energies and
structural features (polarization, FeO$_6$ octahedral tilts and
strains) of a small set of relevant polymorphs. We have applied the
proposed approach to BiFeO$_3$ and La$_{0.25}$Bi$_{0.75}$FeO$_3$,
showing its overall accuracy in reproducing the DFT data. We have also
showed that our models can be used -- by interpolation -- to predict
the properties of compounds with intermediate dopant
concentrations. We note that the introduced potential, as well as the
analytical scheme to obtain its coefficients from DFT, can be readily
applied to study the properties of other perovskite oxides
characterized by the same order parameters (polarization, antiphase
oxygen-octahedral tilts, strains). This includes ferroelectrics where
the tilts are not important (e.g., BaTiO$_{3}$ or PbTiO$_{3}$),
antiferrodistortive non-polar perovskites (e.g., LaAlO$_{3}$), or
compounds where both distortions play a relevant role (e.g.,
SrTiO$_{3}$), as well as their corresponding solid solutions. In
principle, an extension of our scheme to compounds where other order
parameters are relevant (e.g., in-phase tilts in orthorhombic
perovskites like CaTiO$_{3}$ \cite{chen2018})
should be straightforward.

\section{Acknowledgements}
We thank John M. Mangeri for the fruitful discussions.
Work funded by the the Semiconductor Research Corporation and Intel via contract no. 2018-IN-2865. We also acknowledge the support of the Luxembourg National Research Fund through Grant FNR/C18/MS/12705883/REFOX/Gonzalez.

\bibliographystyle{apsrev}
\bibliography{main.bib}
\end{document}